\documentclass[12pt]{article}

\usepackage[top=35mm, bottom=25mm, left=20mm, right=20mm]{geometry}
\usepackage{url, graphicx, color, varioref, amsfonts, longtable, float}
\newcommand{\qed}{\nobreak \ifvmode \relax \else
      \ifdim\lastskip<1.5em \hskip-\lastskip
      \hskip1.5em plus0em minus0.5em \fi \nobreak
      \vrule height0.75em width0.5em depth0.25em\fi}

\title{Constant-Time Quantum Algorithm For The Unstructured Search Problem}

\author{Ahmed Younes\footnote {ayounes2@yahoo.com}\\
Alexandria University\\
Alexandria, Egypt}

\begin{document}
\maketitle
\begin{abstract}
Given an item and a list of values of size $N$. It is required to decide if such item exists in the list. 
Classical computer can search for the item in $O(N)$. The best known quantum algorithm can do 
the job in $O(\sqrt{N})$. In this paper, a quantum algorithm will be proposed that can search an 
unstructured list in $O(1)$ to get the YES/NO answer with certainty.
\end{abstract}

\section{Introduction}       

In 1996, Lov Grover \cite{grover96} presented an algorithm that quantum mechanically searches 
an unstructured list assuming that a unique match exists in the list with quadratic speed-up 
over classical algorithms. The unstructured search problem targeted by Grover's original algorithm 
is deviated in the literature to the following four major problems:
\begin{itemize}
    \item Unstructured list with a unique match.
    \item Unstructured list with one or more matches, where the number of matches is known 
    \item Unstructured list with one or more matches, where the number of matches is unknown.
    \item Unstructured list with strictly multiple matches.
\end{itemize}

The efforts done in all the above cases, similar to Grover's original work, used quantum parallelism by 
preparing superposition that represents all the items in the list. The superposition could be uniform 
or arbitrary. The techniques used in most of the cases to amplify the amplitude(s) of the required 
state(s) have been generalized to an amplitude amplification technique that iterates the operation 
$UR_s \left( \phi  \right)U^\dag  R_t \left( \varphi  \right)$, on $U\left| s \right\rangle$ where 
$U$ is unitary operator, 
$R_s \left( \phi  \right) = I - (1 - e^{i\phi } )\left| s \right\rangle \left\langle s \right|$,
$R_t \left( \varphi  \right) = I - (1 - e^{i\varphi } )\left| t \right\rangle \left\langle t \right|$, 
$\left| s \right\rangle$ is the initial state of the system, $\left| t \right\rangle$ represents 
the target state(s) and $I$ is the identity operator.

Grover's original algorithm replaces $U$ be $W$, where $W$ is the Walsh-Hadamard transform, 
prepares the superposition $W\left| 0 \right\rangle$ (uniform superposition) and iterates 
$WR_s \left( \pi  \right)WR_t \left( \pi  \right)$ for $O\left( {\sqrt N } \right)$, 
where $N$ is the size of the list, which was shown be optimal to get the highest 
probability with the minimum number of iterations \cite{Zalka99}, 
such that there is only one match in the search space.
 
In \cite{Grover98a,Jozsa99,Gal00,long01,BK02}, Grover's algorithm is generalized by showing that 
$U$ can be replaced by almost any arbitrary superposition and the phase shifts $\phi$ and $\varphi$ 
can be generalized to deal with the arbitrary superposition and/or to increase the probability of 
success even with a factor increase in the number of iterations to still run in $O(\sqrt{N})$.
These give a larger class of algorithms for amplitude amplification using variable operators 
from which Grover's algorithm was shown to be a special case.

In another direction, work has been done trying to generalize Grover's algorithm with a uniform 
superposition for known number of multiple matches in the search space 
\cite{boyer96,Chen99,Chen00a,Chen00b}, 
where it was shown that the required number of iterations is approximately 
${\pi}/{4}\sqrt {{N}/{M}}$ for small ${M}/{N}$, where $M$ is the number of matches. 
The required number of iterations will increase for $M>{N}/{2}$, i.e. the problem will be harder 
where it might be excepted to be easier \cite{nc00a}. Another work has been done 
for known number of multiple matches with arbitrary superposition and 
phase shifts \cite{Mosca98,Biron98,Brassard00,hoyer00,Li01} where the same problem 
for multiple matches occurs. In \cite{Brassard98,Mosca98,Brassard00}, 
a hybrid algorithm was presented to deal with this problem 
by applying Grover's fixed operators algorithm for ${\pi}/{4}\sqrt {{N}/{M}}$ 
times then apply one more step using specific $\phi$ and $\varphi$ according to the knowledge of 
the number of matches $M$ to get the solution with probability close to certainty. 
Using this algorithm will increase the hardware cost since we have to build one more 
$R_s$ and $R_t$ for each particular $M$. For the sake of practicality, the operators should be fixed for any 
given $M$ and are able to handle the problem with high probability whether or not $M$ is known in advance. 
In \cite{Younes03d,Younes04a}, Younes et al presented an algorithm that exploits entanglement and partial diffusion 
operator to perform the search and can perform in case of either a single match or 
multiple matches where the number of matches is known or not \cite{Younes04a} 
covering the whole possible range, i.e. $1 \le M \le N$. 
Grover described this algorithm as the best quantum search algorithm \cite{Groverbest}. 
It can be shown that we can get the same probability of success of \cite{Younes03d} using amplitude 
amplification with phase shifts $\phi=\varphi=\pi/2$, 
although the amplitude amplification mechanism will be different. 
The mechanism used to manipulate the amplitudes could be useful in many applications, 
for example, superposition preparation and error-correction. In \cite{YounesA}, an algorithm 
with fixed phase shift operators has been proposed to get a result with probability of success $99.6\%$ 
over the range $1 \le M \le N$ in $O\left(\sqrt {{N}/{M}}\right)$ whether the number of matches is known 
or not in advance.

For unknown number of matches, an algorithm for estimating the number of matches 
({\it quantum counting algorithm}) was presented \cite{Brassard98,Mosca98}. 
In \cite{boyer96}, another algorithm was presented to find a match even if the number of matches is unknown 
which will be able to work if $M$ lies within the range $1\le M \le 3N/4$ \cite{Younes04a}.
 
For strictly multiple matches, Younes et al \cite{Younes03c} presented an algorithm which works 
efficiently only in case of {\it multiple matches} within the search space that splits 
the solution states over more states, inverts the sign of half of them (phase shift of -1) 
and keeps the other half unchanged every iteration. This will keep the mean of the amplitudes 
to a minimum for multiple matches. The same result was rediscovered by Grover using amplitude 
amplification with phase shifts $\phi=\varphi=\pi/3$ \cite{grover150501}, in both algorithms 
the behavior will be similar to the classical algorithms in the worst case.

In this paper, using fixed phase shifts, an algorithm that searches an unstructured list 
in constant-time will be proposed. The algorithm takes the required item $x_s$ and a list $L$ 
as inputs and return an answer with certainty of whether such item exists or not in the list with .

The plan of the paper is as follows: Section \ref{Sec1} introduces the unstructured search problem. 
Section \ref{Sec2} explains the basic operators used in the algorithm. Section \ref{Sec3} proposes 
the algorithm with the trace of its operations. The paper ends up with a conclusion in Section 
\ref{Sec4}.

\section{Unstructured Search Problem}
\label{Sec1}

Consider an item $x_s$ and an unstructured list $L$ of $N$ items. For simplicity and without loss of generality 
we will assume that $N = 2^n$ for some positive integer $n$. Suppose the items in the list are 
labeled with the integers $\{0,1,...,N - 1\}$, and consider a function (oracle) $f$ which maps an 
item $x \in L$ to either 0 or 1 according to some properties this item 
should satisfy, i.e. $f:L \to \{ 0,1\}$. The problem is to find if $x_s$ exists in list assuming 
that at most one $x_s$ exists in the list. In conventional computers, 
solving this problem needs $O\left({N}\right)$ calls to the oracle (query).

\section{Basic Operations}
\label{Sec2}

In this section, the basic operations to be used in the algorithm will be explored. 

\subsection{Hadamard Gate}

The Hadamard gate is a pure quantum gate with special importance in setting up the superposition of 
a quantum register during the quantum computation process. 
Applying the Hadamard gate on a qubit in state $\left| 0 \right\rangle$ or $\left| 1 \right\rangle$ will produce 
a qubit in a perfect superposition, i.e. on measuring the qubit, we will get either $\left| 0 \right\rangle$ or 
$\left| 1 \right\rangle$ with equal probabilities. If $H$ is applied twice, the original input 
state is restored (reversibility). Its truth table is shown in Table~(\ref{Tab1}).

\begin{table}[H]
\begin{center}
\begin{tabular}{|c|c|}
\hline
Input  & 
Output  \\
\hline
$\left| 0 \right\rangle $& 
$\frac{1}{\sqrt 2 }(\left| 0 \right\rangle + \left| 1 \right\rangle )$ \\
\hline
$\left| 1 \right\rangle $& 
$\frac{1}{\sqrt 2 }(\left| 0 \right\rangle - \left| 1 \right\rangle )$ \\
\hline
\end{tabular}
\caption{The Hadamard gate truth table.}
%\label{tab5}
\label{Tab1}
\end{center}
\end{table}

Unitary matrix representation,

\begin{equation}
%\label{eq5}
H = \frac{1}{\sqrt 2 }\left[ {{\begin{array}{*{20}c}
 1 \hfill & \,\,\,\,1 \hfill \\
 1 \hfill & { - 1} \hfill \\
\end{array} }} \right].
\end{equation}

The effect of applying $H$ gate on a single qubit can be understood as follows,

\begin{equation}
H\left| x \right\rangle  = \frac{1}{{\sqrt 2 }}\sum\limits_{y \in \{ 0,1\} } {\left( { - 1} \right)^{x.y} \left| y \right\rangle ,} 
\end{equation}

\noindent
where $x.y$ is the bitwise-AND of $x$ and $y$. Applying $H$ twice gives the original State. i.e. 

\begin{equation}
H\left( {\frac{1}{{\sqrt 2 }}\sum\limits_{y \in \{ 0,1\} } {\left( { - 1} \right)^{x.y} \left| y \right\rangle } } \right) = H\left( {H\left| x \right\rangle } \right)= \left| x \right\rangle .
\end{equation}

In general, the effect of applying $H$ gate on $n$-qubit quantum register 
can be understood as follows,

\begin{equation}
H^{ \otimes n} \left| x \right\rangle  = \frac{1}{{\sqrt {2^n } }}\sum\limits_{y = 0}^{2^n  - 1} {\left( { - 1} \right)^{x.y} \left| y \right\rangle ,} 
\end{equation}

\noindent
where $x.y = \sum\limits_{j = 0}^{n - 1} {x_j .y_j }$ is the summation of 
the bitwise-AND of $x_j$ and $y_j$.

\subsection{NOT Gate}

This quantum gate performs similarly to the classical $NOT$ gate. 
It inverts the state $\left| x \right\rangle $ to the state $\left| \overline x 
\right\rangle$, where $x$ is any Boolean variable and $\overline x$ is its negation. 
Its truth table is shown in Table~(\ref{Tab2}).

\begin{table}[H]
\begin{center}
\begin{tabular}{|c|c|}
\hline
Input  & 
Output   \\
\hline
$\left| 0 \right\rangle $& 
$\left| 1 \right\rangle $ \\
\hline
$\left| 1 \right\rangle $& 
$\left| 0 \right\rangle $ \\
\hline
\end{tabular}
\caption{The $NOT$ gate truth table.}
%\label{fliptt}
\label{Tab2}
\end{center}
\end{table}

Unitary matrix representation,

\begin{equation}
%\label{ch2eqn3.3}
\label{ch2eqn2.24}
X = \left[ {{\begin{array}{*{20}c}
 0 \hfill & 1 \hfill \\
 1 \hfill & 0 \hfill \\
\end{array} }} \right].
\end{equation}

\subsection{Phase Shift}

The phase shift operator will be used to apply a phase shift of -1 
on the amplitude of the state $\left| \mbox{1} \right\rangle $ and leaves the amplitude 
of $\left| \mbox{0} \right\rangle $ with no change. 
Its truth table is shown in Table~(\ref{Tab3}).

\begin{table}[H]
\begin{center}
\begin{tabular}
{|c|c|}
\hline
Input& 
Output   \\
\hline
$\left| 0 \right\rangle $& 
$\,\,\,\left| 0 \right\rangle $ \\
\hline
$\left| 1 \right\rangle $& 
$ -\left| 1 \right\rangle $ \\
\hline
\end{tabular}
\caption{The Phase gate truth table.}
%\label{phasestt}
\label{Tab3}
\end{center}
\end{table}

Unitary matrix representation,

\begin{equation}
%\label{phasesum}
Z = \left[ {{\begin{array}{*{20}c}
 1 \hfill & \,\,\,\,0 \hfill \\
 0 \hfill &  -1 \hfill \\
\end{array} }} \right].
\end{equation}

Such operation will be used to apply a phase shift of $-1$ on a subspace 
of the system entangled with state $\left| \mbox{1} \right\rangle $ as follows,

\begin{equation}
\left(I^{ \otimes n}  \otimes Z \right) \left( {\alpha _0 \left| {\psi _0 } \right\rangle  \otimes \left| 0 \right\rangle  + \alpha _1 \left| {\psi _1 } \right\rangle  \otimes \left| 1 \right\rangle } \right) = \left( {\alpha _0 \left| {\psi _0 } \right\rangle  \otimes \left| 0 \right\rangle  - \alpha _1 \left| {\psi _1 } \right\rangle  \otimes \left| 1 \right\rangle } \right),
\end{equation}

\noindent
where $I$ is the identity operator, ${\left| {\psi _0 } \right\rangle }$ and 
${\left| {\psi _1 } \right\rangle }$ are sub-systems entangled with $\left| \mbox{0} \right\rangle$
and $\left| \mbox{1} \right\rangle $ respectively.

\subsection{Marking Items in a Superposition}

In the literature, there are two ways used to mark certain items in a superposition. 
One way is to conditionally apply certain phase shifts on the marked items 
and the other way is to entangle the required items with certain state of an extra 
working qubit. An oracle $U_f$ 
is used in both cases to recognize the items to be marked, where $f$ is a Boolean 
function evaluates to true for the required items. To mark an item using a phase 
shift of $\alpha$, an oracle $U_{f_{\alpha}}$ of the following effect has been used,

\begin{equation}
U_{f_\alpha  } \left| x \right\rangle  = e^{i\alpha f(x)} \left| x \right\rangle,
\end{equation}

\noindent
and to mark an item by entanglement, an oracle $U_{f_{x}}$ of the following effect has been used,

\begin{equation}
U_{f_x } \left| {x,y} \right\rangle  = \left| {x,y \oplus f(x)} \right\rangle .
\end{equation}

In the proposed algorithm, a combination of both methods will be used where an oracle 
of the form $\exp (i\alpha U_f )$ is used, where $U_f$ has the following effect,

\begin{equation}
U_f \left| {x,0} \right\rangle  = \left| {x,f(x)} \right\rangle.
\end{equation}

Using Taylor's expansion, $\exp (i\alpha U_f)$ can be re-written as,

\begin{equation}
\mathop e\nolimits^{i\alpha U_f }  = \cos (\alpha ).I + i\sin (\alpha ).U_f.
\end{equation}

The effect of applying the oracle $\exp (i\alpha U_f )$ on a superposition of $n+1$ qubit register 
can be understood as follows,

\begin{equation}
\begin{array}{l}
 \mathop e\nolimits^{i\alpha U_f } \left( {\frac{1}{{\sqrt {2^n } }}\sum\limits_{x = 0}^{2^n  - 1} {\left| x \right\rangle  \otimes \left| 0 \right\rangle } } \right) = \left( {\cos (\alpha ).I + i\sin (\alpha ).U_f } \right)\left( {\frac{1}{{\sqrt {2^n } }}\sum\limits_{x = 0}^{2^n  - 1} {\left| x \right\rangle  \otimes \left| 0 \right\rangle } } \right) \\ 
 \,\,\,\,\,\,\,\,\,\,\,\,\,\,\,\,\,\,\,\,\,\,\,\,\,\,\,\,\,\,\,\,\,\,\,\,\,\,\,\,\,\,\,\,\,\,\,\,\,\,\,\,\,\,\,\,\,\,\,\,\,\,\,\,\,\,\, = \frac{{\cos \left( \alpha  \right)}}{{\sqrt {2^n } }}\sum\limits_{x = 0}^{2^n  - 1} {\left| x \right\rangle  \otimes \left| 0 \right\rangle }  + \frac{{i\sin (\alpha )}}{{\sqrt {2^n } }}\sum\limits_{x = 0}^{2^n  - 1} {\left| x \right\rangle  \otimes \left| {f(x)} \right\rangle } . \\ 
 \end{array}
\end{equation}

\subsection{Square Root of NOT with a Global Phase Shift}

The $H_i$ gate is a pure quantum gate. Applying the $H_i$ gate on a qubit in state $\left| 0 \right\rangle$ or 
$\left| 1 \right\rangle$ will produce a qubit in a perfect superposition with some phase shift.
Applying $H_i$ gate twice produces the negation of the original input with some global phase shift. 
Its truth table is shown in Table~(\ref{Tab4}).

\begin{table}[H]
\begin{center}
\begin{tabular}{|c|c|}
\hline
Input  & 
Output  \\
\hline
$\left| 0 \right\rangle $& 
$\frac{1}{\sqrt 2 }(i\left| 0 \right\rangle + \left| 1 \right\rangle )$ \\
\hline
$\left| 1 \right\rangle $& 
$\frac{1}{\sqrt 2 }(\left| 0 \right\rangle + i\left| 1 \right\rangle )$ \\
\hline
\end{tabular}
\caption{The $H_i$ gate truth table.}
%\label{tab5}
\label{Tab4}
\end{center}
\end{table}

Unitary matrix representation,

\begin{equation}
%\label{eq5}
H_i = \frac{1}{\sqrt 2 }\left[ {{\begin{array}{*{20}c}
 i \hfill & 1 \hfill \\
 1 \hfill & {i} \hfill \\
\end{array} }} \right].
\end{equation}

The effect of applying $H_i$ gate on a single qubit can be understood as follows,

\begin{equation}
H_i\left| x \right\rangle  = \frac{1}{{\sqrt 2 }}\sum\limits_{y \in \{ 0,1\} }  { e^{i\frac{\pi}{2}\left({\overline x  \oplus y}\right)}}  \left| y \right\rangle, 
\end{equation}

\noindent
where $x\oplus y$ is the bitwise-XOR of $x$ and $y$, and $\overline x = x \oplus 1$. Applying $H_i$ twice gives the following, 

\begin{equation}
H_i \left( \frac{1}{{\sqrt 2 }}\sum\limits_{y \in \{ 0,1\} } { { e^{i\frac{\pi}{2}\left(\overline x  \oplus y\right)}}  \left| y \right\rangle}\right) = e^{i\frac{\pi}{2}} \left| \overline x \right\rangle.
\end{equation}

In general , the effect of applying $H_i$ gate on $n$-qubit quantum register 
can be understood as follows,

\begin{equation}
H_i^{ \otimes n} \left| x \right\rangle  = \frac{1}{{\sqrt {2^n } }}\sum\limits_{y = 0}^{2^n  - 1} { { e} ^{i\frac{\pi}{2}\left(\overline x\oplus y\right)} \left| y \right\rangle ,} 
\end{equation}

\noindent
where $x\oplus y = \sum\limits_{j = 0}^{n - 1} {x_j \oplus y_j }$ is the summation of 
the bitwise-XOR of $x_j$ and $y_j$. Applying $H_i^{ \otimes n}$ twice gives,
\begin{equation}
H_i^{ \otimes n}\left(\frac{1}{{\sqrt {2^n } }}\sum\limits_{y = 0}^{2^n  - 1} { { e} ^{i\frac{\pi}{2}\left(\overline x\oplus y\right)} \left| y \right\rangle}\right)=e^{i\frac{\pi}{2}n} \left| \overline x \right\rangle.
\end{equation}

\subsection{Phase Shifts Based on Hamming Distance}

The operator $U_c ^{\left| {x_s } \right\rangle }$ is an operator that applies specific phase shifts on the states included in the superposition 
based on the Hamming distance between these states and the given item $x_s$. 
The operator $U_c ^{\left| {x_s } \right\rangle }$ applies phase shifts according to the following rule,

\begin{equation}
U_c ^{\left| {x_s } \right\rangle } \left| x \right\rangle  = \left\{ 
{\begin{array}{*{20}l}
   {e^{i.0} ,} &\mbox{if\,\,} {D(x,x_s ) = 0\,\,\mbox{or}\,\,4n - 3,}  \\
   {e^{i{\textstyle{\pi  \over 2}}} ,} &\mbox{if\,\,} {D(x,x_s ) = 4n - 2,}  \\
   {e^{i\pi } ,} &\mbox{if\,\,} {D(x,x_s ) = 4n - 1,}  \\
   {e^{i{\textstyle{{3\pi } \over 2}}} ,} &\mbox{if\,\,} {D(x,x_s ) = 4n,}  \\
\end{array}} \right.
\end{equation}

\noindent
where $n=1,2,3,...$.

\begin{table}[h]
\begin{center}
\begin{tabular}{|c|c|c|c|c|c|c|c|c|}
  \hline
   & $\left| {000} \right\rangle$ & $\left| {001} \right\rangle$ & $\left| {010} \right\rangle$ & $\left| {011} \right\rangle$ & $\left| {100} \right\rangle$ & $\left| {101} \right\rangle$ & $\left| {110} \right\rangle$ & $\left| {111} \right\rangle$ \\\hline
  $\left| {000} \right\rangle$ &1 &1 &1 &$i$ &1 &$i$ &$i$ &-1\\\hline
  $\left| {001} \right\rangle$ &1 &1 &$i$ &1 &$i$ &1 &-1&$i$ \\\hline
  $\left| {010} \right\rangle$ &1 &$i$ &1 &1 &$i$ &-1&1 &$i$ \\\hline
  $\left| {011} \right\rangle$ &$i$ &1 &1 &1 &-1&$i$ &$i$ &1 \\\hline
  $\left| {100} \right\rangle$ &1 &$i$ &$i$ &-1&1 &1 &1 &$i$ \\\hline
  $\left| {101} \right\rangle$ &$i$ &1 &-1&$i$ &1 &1 &$i$ &1 \\\hline
  $\left| {110} \right\rangle$ &$i$ &-1&1 &$i$ &1 &$i$ &1 &1 \\\hline
  $\left| {111} \right\rangle$ &-1&$i$ &$i$ &1 &$i$ &1 &1 &1 \\
  \hline
\end{tabular}
\caption{Table of phase shifts based on Hamming Distance for 3-qubit states.}
%\label{pi8tt}
\label{TabH}
\end{center}
\end{table}

To construct such operator, for a given $x_s$, choose the corresponding row/column for that item 
from Table~(\ref{TabH}) and insert these values as the diagonal of zero elements matrix. For example, 
if $x_s  = 111$, then the corresponding matrix is, 

\begin{equation}
U_c ^{\left| {111} \right\rangle }  = \left[ {\begin{array}{*{20}c}
   { - 1} & 0 & 0 & 0 & 0 & 0 & 0 & 0  \\
   0 & i & 0 & 0 & 0 & 0 & 0 & 0  \\
   0 & 0 & i & 0 & 0 & 0 & 0 & 0  \\
   0 & 0 & 0 & 1 & 0 & 0 & 0 & 0  \\
   0 & 0 & 0 & 0 & i & 0 & 0 & 0  \\
   0 & 0 & 0 & 0 & 0 & 1 & 0 & 0  \\
   0 & 0 & 0 & 0 & 0 & 0 & 1 & 0  \\
   0 & 0 & 0 & 0 & 0 & 0 & 0 & 1  \\
\end{array}} \right]
\end{equation}

To simplify the construction of $U_c ^{\left| {x_s } \right\rangle }$, instead of 
choosing the appropriate row/column from Table~(\ref{TabH}). The same construction can be done 
as follows,

\begin{equation}
U_c ^{\left| {x_s } \right\rangle }  = X^{ \otimes \neg \left\langle {x_s } \right\rangle } U_c ^{\left| 1 \right\rangle ^{ \otimes n} } X^{ \otimes \neg \left\langle {x_s } \right\rangle },
\end{equation}
 
\noindent
where ${\left\langle {x_s } \right\rangle }$ is the bit representation of $x_s$, and $\neg$ is the 
bitwise negation operator. For example, if $x_s  = 101$, then,

\begin{equation}
U_c ^{\left| {101} \right\rangle }  = \left( {I \otimes X \otimes I} \right)U_c ^{\left| {111} \right\rangle }\left( {I \otimes X \otimes I} \right).
\end{equation}

\section{The Algorithm}
\label{Sec3}

Given a list $L$ of size $N=2^n$ and an item $x_s$. It is required to decide if $x_s$ is in 
the list. The operations of the algorithm is applied as follows,

\begin{equation}
\left( {H_i ^{ \otimes n}  \otimes I} \right)\left( {U_c ^{\left| {x_s } \right\rangle }  \otimes I} \right) e^{i{\textstyle{\pi  \over 4}}U_f } \left( {I^{ \otimes n}  \otimes Z} \right)e^{i{\textstyle{\pi  \over 4}}U_f } \left( {H^{ \otimes n}  \otimes I} \right)\left| 0 \right\rangle ^{ \otimes n+1}.  \end{equation}

\subsection{Tracing the Algorithm}

The steps of the algorithm are as follows:

\begin{itemize}

\item[1-] Prepare a quantum register of size $n+1$ qubits all in state $\left| 0 \right\rangle$.

\begin{equation}
\left| {\psi _0 } \right\rangle  = \left| 0 \right\rangle ^{ \otimes n}  \otimes \left| 0 \right\rangle .
\end{equation}

\item[2-] Apply $H$ gate on each of the first $n$ qubits.

\begin{equation}
\begin{array}{l}
 \left| {\psi _1 } \right\rangle  = \left( {H^{ \otimes n}  \otimes I} \right)\left| {\psi _0 } \right\rangle  \\ 
 \,\,\,\,\,\,\,\,\,\,\,\, = \frac{1}{{\sqrt {2^n } }}\sum\limits_{x = 0}^{2^n  - 1} {\left| x \right\rangle  \otimes \left| 0 \right\rangle } . \\ 
 \end{array}
\end{equation}

\item[3-] Apply $\exp (i\alpha U_f )$ taking $\alpha  = \frac{\pi }{4}$.

\begin{equation}
\begin{array}{l}
 \left| {\psi _2 } \right\rangle  = \exp (i\frac{\pi }{4}U_f )\left| {\psi _1 } \right\rangle  \\ 
 \,\,\,\,\,\,\,\,\,\, = \left( {\frac{1}{{\sqrt 2 }}.I + \frac{i}{{\sqrt 2 }}.U_f } \right)\frac{1}{{\sqrt {2^n } }}\sum\limits_{x = 0}^{2^n  - 1} {\left| x \right\rangle  \otimes \left| 0 \right\rangle }  \\ 
 \,\,\,\,\,\,\,\,\,\, = \frac{1}{{\sqrt {2^{n + 1} } }}\left( {\sum\limits_{x = 0}^{2^n  - 1} {\left( {\left| x \right\rangle  \otimes \left| 0 \right\rangle } \right)}  + i\sum\limits_{x = 0}^{2^n  - 1} {\left( {\left| x \right\rangle  \otimes \left| {f(x)} \right\rangle } \right)} } \right). \\ 
 \end{array}
\end{equation}

If $x_s$ exists in the list, then the system can be written as,

\begin{equation}
\left| {\psi _2 } \right\rangle  = \frac{{i + 1}}{{\sqrt {2^{n + 1} } }}\sum\limits_{\scriptstyle x = 0 \hfill \atop 
  \scriptstyle x \ne x_s  \hfill}^{2^n  - 1} {\left( {\left| x \right\rangle  \otimes \left| 0 \right\rangle } \right)}  + \frac{1}{{\sqrt {2^{n + 1} } }}\left| {x_s } \right\rangle  \otimes \left( {\left| 0 \right\rangle  + i\left| 1 \right\rangle } \right).
\end{equation}

\item[4-] Apply $\left( {I^{ \otimes n}  \otimes Z} \right)$.

\begin{equation}
\begin{array}{l}
 \left| {\psi _3 } \right\rangle  = \left( {I^{ \otimes n}  \otimes Z} \right) \left| {\psi _2 } \right\rangle  \\ 
 \,\,\,\,\,\,\,\,\,\,\,\, = \frac{{i + 1}}{{\sqrt {2^{n + 1} } }}\sum\limits_{\scriptstyle x = 0 \hfill \atop 
  \scriptstyle x \ne x_s  \hfill}^{2^n  - 1} {\left( {\left| x \right\rangle  \otimes \left| 0 \right\rangle } \right)}  + \frac{1}{{\sqrt {2^{n + 1} } }}\left| {x_s } \right\rangle  \otimes \left( {\left| 0 \right\rangle  - i\left| 1 \right\rangle } \right). \\ 
 \end{array}
\end{equation}

\item[5-] Apply $\exp (i\alpha U_f )$ taking $\alpha  = \frac{\pi }{4}$.

\begin{equation}
\begin{array}{l}
 \left| {\psi _4 } \right\rangle  = \exp (i\frac{\pi }{4}U_f )\left| {\psi _3 } \right\rangle  \\ 
 \,\,\,\,\,\,\,\,\,\,\,\, = \frac{i}{{\sqrt {2^n } }}\sum\limits_{\scriptstyle x = 0 \hfill \atop 
  \scriptstyle x \ne x_s  \hfill}^{2^n  - 1} {\left( {\left| x \right\rangle  \otimes \left| 0 \right\rangle } \right)}  + \frac{1}{{\sqrt {2^n } }}\left| {x_s } \right\rangle  \otimes \left| 0 \right\rangle.  \\ 
 \end{array}
\end{equation}

\item[6-] Apply $\left( {U_c ^{\left| {x_s } \right\rangle }  \otimes I} \right)$.

\begin{equation}
\begin{array}{l}
 \left| {\psi _5 } \right\rangle  = \left( {U_c ^{\left| {x_s } \right\rangle }  \otimes I} \right) \left| {\psi _4 } \right\rangle  \\ 
 \,\,\,\,\,\,\,\,\,\,\,\, = \frac{1}{{\sqrt {2^n } }}\sum\limits_{\scriptstyle x = 0 \hfill \atop 
  \scriptstyle x \ne x_s  \hfill}^{2^n  - 1} {e^{i\frac{{m\pi }}{2}} \left( {\left| x \right\rangle  \otimes \left| 0 \right\rangle } \right)}  + \frac{1}{{\sqrt {2^n } }}\left| {x_s } \right\rangle  \otimes \left| 0 \right\rangle,\\ 
 \end{array}
\end{equation}
\noindent
where $m= x_s  \oplus x = \sum\limits_{j = 0}^{n - 1} {x_{s_j }  \oplus x_j }  = 1,2,3,\ldots$. The system can be re-written as,

\begin{equation}
\left| {\psi _5 } \right\rangle \,\,\, = \frac{1}{{\sqrt {2^n } }}\sum\limits_{x = 0}^{2^n  - 1} {e^{i\frac{{m_s \pi }}{2}} \left( {\left| x \right\rangle  \otimes \left| 0 \right\rangle } \right),} \,\,\,\,\,\,\,m_s  = 0,1,2,3,...
\end{equation}

\item[7-] Apply $H_i$ gate on each of the first $n$ qubits.

\begin{equation}
\left| {\psi _6 } \right\rangle  = \left( {H_i ^{ \otimes n}  \otimes I} \right)\left| {\psi _5 } \right\rangle  = e^{i\frac{{n\pi }}{2}} \left| {x_s } \right\rangle  \otimes \left| 0 \right\rangle .
\end{equation}

\item[8-]Measure the first $n$ qubits. If the outcome is $x_s$, then the required item exists in the 
list, otherwise, the item doesn't exist.
%\begin{itemize}
%	\item[i-]  If the result is an item $x \ne x_s$, then $x_s$ doesn't exist in the system.
%	\item[ii-] If the result is the item $x_s$, then the item might exist in the list. 
%	This arises as an effect of applying $\left( {H_i ^{ \otimes n}  \otimes I}\right)$. If the item exists 
%	then the system will contain $x_s$ with certainity. 
%	If the item doesn't exist then applying $\left( {H_i ^{ \otimes n}  \otimes I}\right)$ will create 
%	a superposition of items with $x_s$ is included among them. In this case, 
%	applying a measurement, we might get $x_s$ with probabilty $p=1 - \frac{1}{{2^n }}$. 
%	Applying the algorithm 1/p times, we can decide if the item $x_s$ does exist in the system. 
%	For suffient large size of $L$, applying the algorithm one or two times might be suffienet 
%	to get the answer.
%\end{itemize}
\end{itemize}

\section{Conclusion}
\label{Sec4}

Using quantum superposition and fixed phase shifts, a quantum computer can search an unstructured 
list in a single step. The algorithm used a phase shift and a temporary entangelemnt to mark the 
item within the search space. An operator is used to adjust the phases of the items 
in the list according to their Hamming distance with the required item. Finally, we get an 
answer with certainty of whether the item exists or not in the list.

\bibliography{O_1}
\bibliographystyle{plain}

\end{document}